# KIC 3868420: A High-amplitude $\delta$ Scuti–$\gamma$ Dor Hybrid Star Crossing the Hertzsprung Gap

Tao-Zhi Yang and Zhao-Yu Zuo
Ministry of Education Key Laboratory for Nonequilibrium Synthesis and Modulation of Condensed Matter, School of Physics, Xi'an Jiaotong University, Xi'an 710049, People's Republic of China; yangtaozhi2018@163.com, zuozyu@xjtu.edu.cn

## Abstract

We report a photometric and asteroseismic analysis of KIC 3868420, a newly identified high-amplitude $\delta$ Scuti–$\gamma$ Doradus hybrid star located in the Hertzsprung gap—a short-lived and rarely observed post-main-sequence phase. Using 4 yr of Kepler long-cadence photometry, we detect 36 significant frequencies, including 11 independent modes spanning both low- and high-frequency regimes. Grid-based modeling with MESA and GYRE, including rotation, shows that five independent frequencies match a combination of radial $p$-modes and nonradial $g$-modes, supporting its hybrid nature. The best-fit models yield an evolved post-main-sequence star ($M \sim 2.26$–$2.30 M_\odot$, $R \sim 4.41$–$4.43 R_\odot$, and $\tau \sim 5.4 \times 10^8$ yr), although degeneracies from rotation and unknown inclination preclude a unique solution. KIC 3868420 thus represents a rare example of a high-amplitude hybrid pulsator in the Hertzsprung gap, demonstrating that high-amplitude pulsation can coexist with mixed $p$-/$g$-mode behavior in a rapidly evolving star. This finding highlights the value of space-based photometry for identifying such rare objects.

*Unified Astronomy Thesaurus concepts:* Stellar oscillations (1617); Delta Scuti variable stars (370)

## 1. Introduction

Understanding the internal structures and evolutionary processes of stars is a fundamental goal of stellar astrophysics. While observations across the electromagnetic spectrum have long provided tight constraints on stellar atmospheres and global parameters, direct insight into stellar interiors remains elusive (A. Maeder & G. Meynet 2000; A. Maeder 2009). Consequently, theoretical models must be tested with diagnostic tools capable of probing beneath the stellar surface. Asteroseismology, the study of stellar oscillations, provides such an important tool for this purpose. Oscillation frequencies are highly sensitive to internal conditions; their measurement and interpretation offer a powerful means to test models of stellar structure and evolution (C. Aerts et al. 2010; C. Aerts 2021). By comparing observed oscillation spectra with theoretical models, fundamental stellar parameters such as mass, radius, and age can be derived. Furthermore, asteroseismology can provide constraints on internal processes, including convection (L. Grassitelli et al. 2015; W. D. Arnett et al. 2019), mixing (M. G. Pedersen et al. 2021), rotation (G. Li et al. 2019, 2020), and magnetic fields (J. O. Sundqvist et al. 2013; L. Bugnet 2022) that critically influence stellar evolutionary paths but are difficult to constrain from classical observables alone.

$\delta$ Scuti ($\delta$ Sct) stars are intermediate-mass (1.5–2.5 $M_\odot$) A- and F-type pulsators within the classical instability strip, exhibiting low-order pressure ($p$) and gravity ($g$) modes driven by the $\kappa$ mechanism in the He II ionization zone (M. Breger 2000; C. Aerts et al. 2010; D. M. Bowman & D. W. Kurtz 2018). They are located in the lower part of the classical instability strip and span the main sequence in the Hertzsprung–Russell (H-R) diagram. Stars near the zero-age main sequence (ZAMS) typically have effective temperatures between 7500 and 9500 K, while those near the terminal-age main sequence (TAMS) range from about 6500 to 8500 K (D. M. Bowman & D. W. Kurtz 2018; S. J. Murphy et al. 2019). The presence of both radial and nonradial $p$-mode pulsations in these stars enables probing of the stellar envelope. While their traditionally complex spectra posed challenges for mode identification, space-based photometry from missions like CoRoT and Kepler has revolutionized the field by enabling pattern-based diagnostics. Key advances include the identification of regular, large-separation-like frequency spacings that provide tools for mode identification and mean-density constraints (A. García Hernández et al. 2013), theoretical calibrations of these spacings showing resilience to moderate rotation (J. C. Suárez et al. 2014; J. E. Rodríguez-Martín et al. 2020), the establishment of empirical scaling relations from ensemble studies (J. E. Rodríguez-Martín et al. 2020), and the quantitative characterization of the low-amplitude "grass" in frequency spectra as a seismic diagnostic (S. Barceló Forteza et al. 2024), collectively motivating a more standardized approach to $\delta$ Sct asteroseismology.

Adjacent to $\delta$ Sct stars in the H-R diagram lie $\gamma$ Doradus ($\gamma$ Dor) stars. These stars pulsate in high-order, low-degree $g$-modes with periods typically between 0.3 and 3 days. These modes are sensitive to the deep interior, especially the near-core region, providing valuable insight into internal stellar processes. Importantly, the asymptotic theory for high-order $g$-modes predicts nearly uniform period spacings (W. Unno et al. 1979). These period spacings and their deviations have been used to study internal mixing processes, rotation, and chemical composition gradients (A. Miglio et al. 2008; T. Van Reeth et al. 2015, 2016, 2018; G. Li et al. 2019, 2020). The instability strips of $\delta$ Sct and $\gamma$ Dor stars overlap in the H-R diagram (L. A. Balona 2011; G. W. Henry et al. 2011;

K. Uytterhoeven et al. 2011; D. R. Xiong et al. 2016), giving rise to hybrid pulsators that exhibit both high-frequency $p$-mode pulsations and low-frequency $g$-mode oscillations. The first such hybrid was discovered by G. W. Henry & F. C. Fekel (2005), followed by others such as HD 49434 and HD 8801 (K. Uytterhoeven et al. 2008; G. Handler 2009).

The overlapping instability strips of $\delta$ Sct and $\gamma$ Dor stars give rise to hybrid pulsators that exhibit both high-frequency $p$-modes and low-frequency $g$-modes, offering a simultaneous probe of the stellar envelope and core (A. Grigahcène et al. 2010; K. Uytterhoeven et al. 2011). Recent findings have significantly revised the classical view of high-amplitude $\delta$ Sct (HADS) stars. It is now established that large amplitudes do not preclude rich frequency spectra containing nonradial modes (C. Lv et al. 2022). Crucially, analyses of space-based data reveal that many HADS are likely evolved stars, occupying the cooler side of the instability strip and showing signatures of mixed $p$–$g$ pulsation behavior that persists into the post-main-sequence phase, thereby blurring the distinction between classical HADS and hybrid pulsators (C. Lv et al. 2023; F. Vasigh et al. 2024; E. Ziaali et al. 2025). This establishes evolved, high-amplitude hybrids as an observationally emergent class of objects. Accurately modeling such potentially evolved A-/F-type pulsators and deriving their fundamental parameters necessitates a careful treatment of rotational effects.

Rotation affects $\delta$ Sct asteroseismology—and is particularly important for HADS—through two coupled channels (J. C. Suárez et al. 2006a, 2006b, 2007): (1) *structural modification*, where the centrifugal contribution reduces the effective gravity and alters the equilibrium stratification (often via pseudo-rotating models), shifting zeroth-order frequencies; and (2) *mode-dependent perturbations*, where second-order rotation and near-degeneracy reshape the spectrum, with mixed/$g$-like modes additionally sensitive to the internal rotation profile, producing $\mu$Hz-level shifts across both the low-frequency (mixed/$g$-) and $p$-mode regimes (J. C. Suárez et al. 2007). Even modest rotation ($\sim$15–50 km s$^{-1}$) can change the Petersen-diagram period ratio $\Pi_{1/0}$ at the $\sim 10^{-2}$ level, creating a strong degeneracy with metallicity in double-mode HADS unless rotation is independently constrained (J. C. Suárez et al. 2006a, 2006b). The first overtone is generally more rotation-sensitive than the fundamental radial mode, and near-degeneracy can introduce additional "wriggles" in $\Pi_{1/0}(\Omega)$ by coupling to $\ell = 2$ modes, while the fundamental remains comparatively stable (J. C. Suárez et al. 2006a, 2006b).

Hybrid stars, particularly large-amplitude $\delta$ Sct–$\gamma$ Dor variables, are extremely valuable targets for studying the effects of rotation on stellar internal structure. KIC 3868420 (HD 225525; $\alpha_{J2000} = 19^h43^m26^s.40$, $\delta_{J2000} = +38°55'12''.46$) was initially thought to be an RRc variable star in early Kepler photometry, but subsequent observations suggested it might be a HADS candidate (J. M. Nemec et al. 2013). Further high-resolution spectroscopic observations indicated that the star was tentatively classified as a potential $\delta$ Sct–$\gamma$ Dor hybrid variable star, so its specific classification remains unclear (E. Niemczura et al. 2017). Table 1 lists some basic parameters of KIC 3868420.

To further investigate its pulsational nature and internal structure, we conduct a detailed asteroseismic analysis of KIC 3868420 mainly based on the 4 yr long-cadence photometric data from the Kepler mission. In this work, we analyze its frequency spectrum, identify pulsation modes, and construct a series of stellar models, including rotation, to derive the star's fundamental parameters and evolutionary state.

**Table 1**
Basic Parameters of KIC 3868420

| Parameters | Values | References |
|---|---|---|
| $K_P$ | 10.110 | (a) |
| $T_{mag}$ | 9.972 | (b) |
| $G$ | 10.246 | (c) |
| $G_{bp}$ | 10.416 | (c) |
| $G_{rp}$ | 9.886 | (c) |
| $B$ | 10.678 | (a) |
| $V$ | 10.310 | (a) |
| $J_{2MASS}$ | 9.172 | (a) |
| $H_{2MASS}$ | 8.918 | (a) |
| $K_{2MASS}$ | 8.790 | (a) |
| $E(B-V)$ | $0.0724 \pm 0.0097$ | (a) |
| $A_V$ | 0.185 | (a) |
| $\pi$ (mas) | $1.045 \pm 0.023$ | (c) |
| $d$ (pc) | $932.063 \pm 20.097$ | (d) |
| $T_{eff}$ (K) | $7480 \pm 150$ | (e) |
| $T_{eff}$ (K) | $7800 \pm 100$ | (f) |
| $\log g$ (dex) | $3.9 \pm 0.1$ | (f) |
| $v \sin i$ (km s$^{-1}$) | $5.7 \pm 1.2$ | (e) |
| [Fe/H] (dex) | $-0.32 \pm 0.13$ | (e) |

**References.** (a) Kepler Input Catalog (KIC; T. M. Brown et al. 2011); (b) TESS Input Catalog (TIC; K. G. Stassun et al. 2019); (c) Gaia Collaboration (2018); (d) C. A. L. Bailer-Jones et al. (2018); (e) J. M. Nemec et al. (2013); (f) E. Niemczura et al. (2017).

## 2. Observation and Data Reduction

KIC 3868420 was observed by Kepler space telescope from BJD 2454953.54 to 2456424.00 days, including 15 long-cadence (LC; 29.5 minute integration time) data: Q0–Q5, Q7–Q9, Q11–Q13, and Q15–Q17, which spans about 1470.46 days. There was also a piece of short-cadence (SC; 59 s integration time) data of 9.7 days (Q0). All the LC and SC data are available in Kepler Asteroseismic Science Operations Center (KASOC) database[1] (H. Kjeldsen et al. 2010). Both the LC and SC time-series data are presented in two types: "raw" and "corrected." The former is from the NASA Kepler Science pipeline, while the latter has been reduced by KASOC Working Group 4 (WG4: $\delta$ Scuti targets). In this work, we downloaded all the LC and SC time-series data of this star and used the corrected flux. For each quarter of LC data, the obvious outliers were removed first, and the slow trend was corrected with a linear or second-order polynomial. Then the flux data was converted to magnitude scale, and each quarter was adjusted to zero by subtracting its mean value. Finally, all the quarters were stitched to a total light curve with 51,855 data points. Figure 1 shows a sample of SC and LC light curves with the same time period, both indicating that KIC 3868420 is a high-amplitude pulsating variable star.

Figure 2 shows the phased light curve, with a peak-to-peak amplitude of approximately 0.15 mag, folded by the dominant frequency $f_1 = 4.8024429$ day$^{-1}$. In contrast to typical HADS stars, the light curve of KIC 3868420 is highly symmetric and distinctly lacks the characteristic rapid rise followed by a slow decline.

---

[1] KASOC database: https://kasoc.phys.au.dk/.

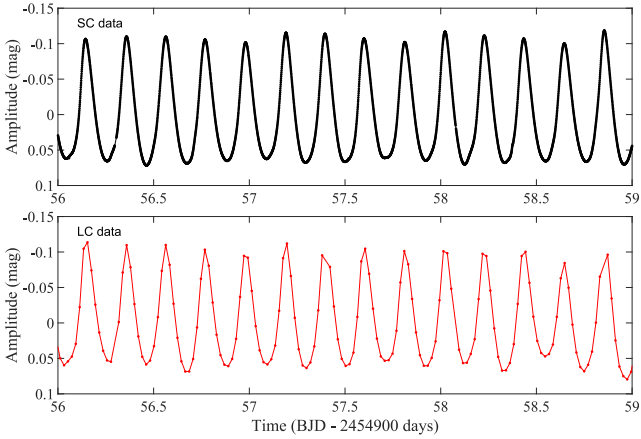

**Figure 1.** Sample light curves of KIC 3868420 from Kepler SC and LC data. Top panel: light curve from SC data. Bottom panel: light curve from LC data. The amplitude of the light curves from both data is larger than 0.15 mag, indicating it is a high-amplitude pulsating star.

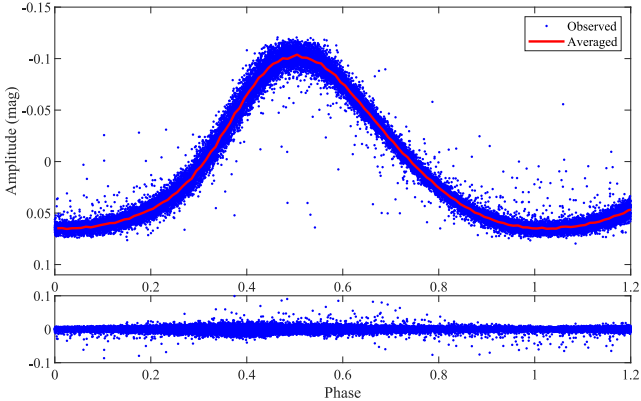

**Figure 2.** Phase diagram of KIC 3868420 from LC data, folded by the dominant frequency, $f_1 = 4.8024429$ day$^{-1}$. The red line shows the average value of the light curve in a bin of 0.01 phase.

## 3. Frequency Analysis and Mode Identification

To investigate the pulsating behavior of KIC 3868420, we extracted the pulsation frequencies with software PERIOD04 (P. Lenz & M. Breger 2005), in which the light curve is fitted with the following formula:

$$m = m_0 + \Sigma A_i \sin(2\pi(f_i t + \phi_i)), \quad (1)$$

where $m_0$ is the zero-point, and $A_i$, $f_i$, and $\phi_i$ are the amplitude, frequency, and phase of the $i$th peak, respectively.

Since the SC data spanned less than 10 days and was insufficient for high-precision work, we therefore limited our frequency extraction and identification to the more extensive LC dataset. To detect more potential pulsation modes, we chose a wider frequency range of $0 < f < 50$ day$^{-1}$, which covers the typical pulsation frequency range of $\delta$ Sct stars. We detected the significant peaks one by one via the standard method of prewhitening as adopted by T.-Z. Yang & A. Esamdin (2019). As the Nyquist frequency of LC data is $f_N = 24.468$ day$^{-1}$ (S. J. Murphy et al. 2013; D. L. Holdsworth et al. 2014), there would be aliases in frequency spectra with the range of $0 < f < 50$ day$^{-1}$. It is not difficult to judge the alias, as they usually show multiplet structures with a frequency interval of $f_{\rm orb} = 0.00268$ day$^{-1}$. We refer the reader to T. Yang et al. (2018) for further details.

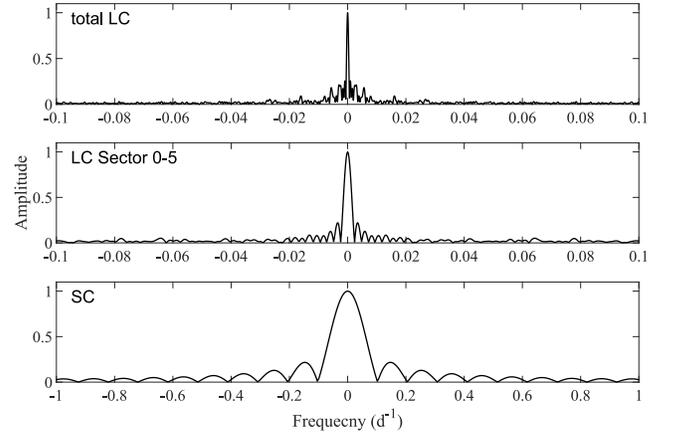

**Figure 3.** Top panel: spectra window of total 4 yr LC data, lacking Q6, Q10, and Q14. Middle panel: spectra window of consecutive sectors 0–5 LC data. Bottom panel: spectra window of one sector SC data (Q0) only spanning 9.7 days. Note that the range of the $x$-axis in the SC panel is 10 times that in the LC panel.

### 3.1. Spectra Window

In asteroseismology of pulsating stars, when performing a Fourier transform on a light curve, it is necessary to simultaneously analyze the spectral window function, as it directly determines what you can "see" in the data, which frequencies are real, and which are just "alias frequencies" caused by observation sampling. Figure 3 shows three spectra windows of different types of data: total 4 yr LC data with a lack of three sectors (Q6, Q10, and Q14), consecutive sectors 0–5 LC data, and 9.7 day SC data. From the figure, it is clear that the spectral window of the total 4 yr light curve has the narrowest center width (top panel), as it has the longest time span. However, due to the interruption of data in three sectors, some visible side lobes appear on both sides of the main peak. Therefore, the total 4 yr light variation curve provides the smallest frequency resolution (Rayleigh frequency = 0.001 day$^{-1}$), but the influence of these side lobes needs to be carefully considered when analyzing and verifying frequencies. For the continuous light curves of sectors 0–5, since its time span is only about one-third of the total light curves mentioned above, the width of its main peak is slightly wider (middle panel), resulting in a smaller frequency resolution (Rayleigh frequency = 0.003 day$^{-1}$). However, due to the continuity of its data, there are no visible side lobes on its main peak. Therefore, these data are crucial for further verification of the authenticity of frequencies obtained from the total LC data. SC data have a very short duration, less than 10 days, resulting in the widest peak (bottom panel) and the weakest frequency resolution (Rayleigh frequency = 0.15 day$^{-1}$). However, due to its high sampling rate (59 s), it has the highest Nyquist frequency (=689.5 day$^{-1}$). Therefore, in frequency ranges exceeding the Nyquist frequency of LC data, SC data can be used to identify alias frequencies.

### 3.2. Real and Alias Frequencies

For Kepler's LC data, its Nyquist frequency is 24.486 day$^{-1}$, which falls within the frequency range of $\delta$

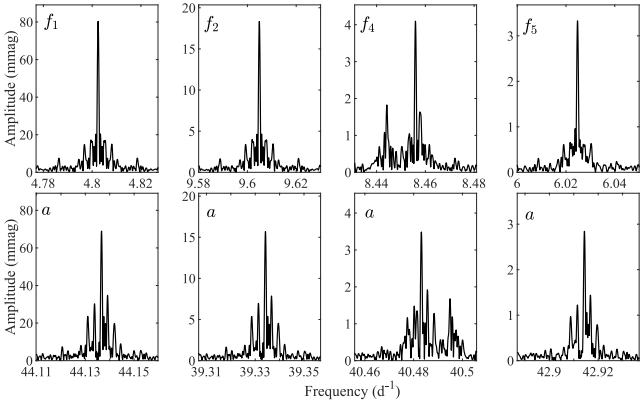

**Figure 4.** Comparison of typical real and alias frequencies in the amplitude spectrum of the 4 yr LC data. Upper panels: four real frequencies. Lower panels: corresponding alias frequencies (labeled by "a") within the super-Nyquist frequency range, which are reflections of the real frequencies.

Scuti variable stars. This means that eigenfrequencies may also exist in the range beyond the Nyquist frequency. S. J. Murphy et al. (2013) proposed a way to distinguish real and super-Nyquist modes in Kepler's light curves thanks to their periodically modulated sampling. Because of the spectral window that this kind of cadence data produces, the super-Nyquist peaks split into a multiplet whose separation is a multiple of the orbital frequency of the Kepler telescope ($f_{\rm orb} = 0.00268\,{\rm day}^{-1}$). For this reason, the amplitude of the peak in the super-Nyquist frequency regime will be lower than the real peak.

Figure 4 shows the amplitude spectra of four real frequencies and their corresponding alias frequencies in the super-Nyquist frequency range. It can be seen that the real frequencies and the spectral window function have the same structure, while the peak of the alias frequencies is not only lower than the real frequencies, but also has distinct sidelobes on both sides, with the frequency interval between the sidelobes equal to the orbital frequency ($=0.00268\,{\rm day}^{-1}$) of the Kepler telescope. These characteristics can be used to effectively identify real and alias frequencies appearing in the LC spectrum.

On the other hand, the high sampling rate of SC data (59 s) causes its Nyquist frequency to reach $f_{\rm Nyq} = 689.5\,{\rm day}^{-1}$. Therefore, comparing the amplitude spectra of LC and SC data can effectively identify the super-Nyquist alias frequencies in LC data. Figures 5 and 6 show a comparison of amplitude spectra of these two types of data in the infra-Nyquist frequency range and the super-Nyquist frequency range, respectively. From Figure 5, it can be seen that for the real frequencies, they can be detected in both LC and SC data. However, due to the longer integration time of LC data, the amplitude of its spectrum is slightly lower than that of SC data. In the super-Nyquist frequency range, as shown in Figure 6, the alias frequencies caused by Nyquist frequency reflections that appear in LC data are absent in SC data. Therefore, in this work, we also utilize SC data for frequency extraction, which helps identify alias frequencies detected in LC.

Figure 7 shows the amplitude spectra and the prewhitening procedures of the LC light curve. For the 4 yr LC data, the window would contain about 3000 independent Rayleigh bins. The real peaks are likely to fall within the region and inflate the mean noise. Therefore, we use a median estimator over a narrower window of 0.5 c/d around the peak to calculate the noise for LC data and 2.0 c/d for SC data. In total, 36 significant frequencies with signal-to-noise ratio (S/N) > 5.4 (A. S. Baran et al. 2015) are extracted in LC data and listed in Table 2. The uncertainties of frequencies were calculated following M. H. Montgomery & D. Odonoghue (1999) and H. Kjeldsen & T. R. Bedding (2012).

Among the 36 significant frequencies, 11 frequencies (i.e., $f_1$, $f_4$, $f_5$, $f_6$, $f_8$, $f_9$, $f_{13}$, $f_{19}$, $f_{20}$, $f_{23}$, and $f_{26}$) are considered as independent frequencies. In high-amplitude pulsating stars, for example, HADS stars, the strongest frequency was usually assumed as the fundamental radial mode, then one can identify the low-order radial modes, according to the relation derived from R. F. Stellingwerf (1979): $0.756 \leqslant P_1/P_0 \leqslant 0.787$, $0.611 \leqslant P_2/P_0 \leqslant 0.632$, and $0.500 \leqslant P_3/P_0 \leqslant 0.525$, where $P_0$, $P_1$, $P_2$, and $P_3$ are the periods of the fundamental, first overtone, second overtone, and third overtone modes, respectively.

For KIC 3868420, we first assumed that $f_1$ belongs to the fundamental mode, then we obtained the ratios as $f_1/f_5 = 0.797$, $f_1/f_4 = 0.568$, and $f_1/f_6 = 0.569$. The ratio 0.797 is inconsistent with the $P_1/P_0$ ratio, and 0.568(0.569) is inconsistent with $P_3/P_0$. This strongly suggests that $f_1$ is not a radial fundamental mode. Other independent frequencies are basically stronger peaks, and they are neither any combinations nor harmonics of other frequencies. They are marked with "independent" in the last column of Table 2.

To identify the combination frequency, we use a simple form: $(f \pm \sigma) = m(f_1 \pm \sigma_1) \pm n(f_2 \pm \sigma_2)$ within the Rayleigh resolution limit ($1/\Delta T = 0.001\,{\rm day}^{-1}$), where $m$ and $n$ are small integers, and $f_1$ and $f_2$ are the stronger frequencies. All the combinations and/or harmonics and their identifications are also listed in Table 2.

### 4. Asteroseismic Modeling

For pulsating stars, accurate asteroseismic models are very helpful to constrain the key fundamental stellar parameters, such as mass, age, and evolutionary status, allowing us to explore the internal structure of stars. As a reward, asteroseismic models can also further verify the mode identification of stellar pulsation frequencies. To determine the fundamental stellar parameters and evolutionary state of KIC 3868420, we performed grid-based forward asteroseismic modeling in two stages: a broad grid search to delimit the parameter space, and subsequent detailed pulsation computations for the most promising models, which included rotational effects to refine both the mode identifications and stellar parameters.

#### 4.1. Stellar Evolution Grid and Preliminary Analysis

We first constructed a preliminary wider grid of stellar evolution tracks with Modules for Experiments in Stellar Astrophysics (MESA v8118; B. Paxton et al. 2018, 2019), varying the stellar mass at the observed metallicity. As J. M. Nemec et al. (2013) reported the spectroscopic iron-to-hydrogen ratio of this star as [Fe/H] = $-0.32 \pm 0.13$, we derived the metallicity $Z = 0.0067^{+0.0023}_{-0.0017}$, according to the relation: [Fe/H] = $\log(Z/X) - \log(Z/X)_\odot$, where the value $(Z/X)_\odot = 0.0181$ (M. Asplund et al. 2009) was adopted. In the preliminary evolutionary grid, the stellar masses were chosen in a range of $2.00-2.40\,M_\odot$ with a step of $0.02\,M_\odot$, and metallicity $Z$ was adopted with five values, namely, from 0.005 to 0.009, with a step of 0.001, covering the range of metal abundance obtained above. For the helium abundance $Y$, we adopted $Y = 0.249 + 1.33 Z$ as a function of $Z$. The classical mixing-length theory with $\alpha = 1.90$ was used in the convective region (E. Böhm-Vitense 1958; B. Paxton et al. 2011).

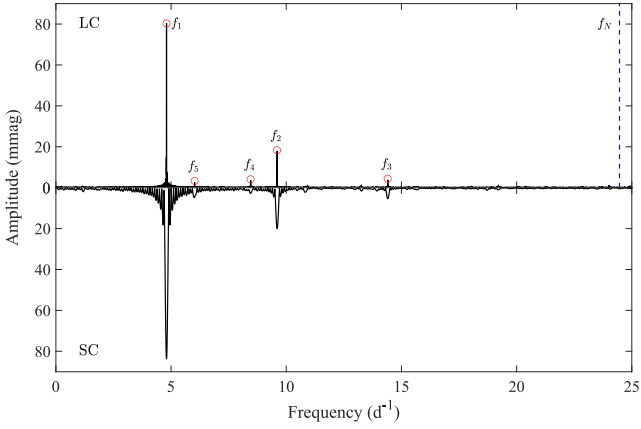

**Figure 5.** Comparison of amplitude spectra of the 4 yr LC and 9.7 day SC data in the infra-Nyquist frequency range. The upper panel shows amplitude spectra of the 4 yr LC; five strong real frequencies are marked by red circles and the inverted lower panel shows amplitude spectra of the 9.7 day SC data.

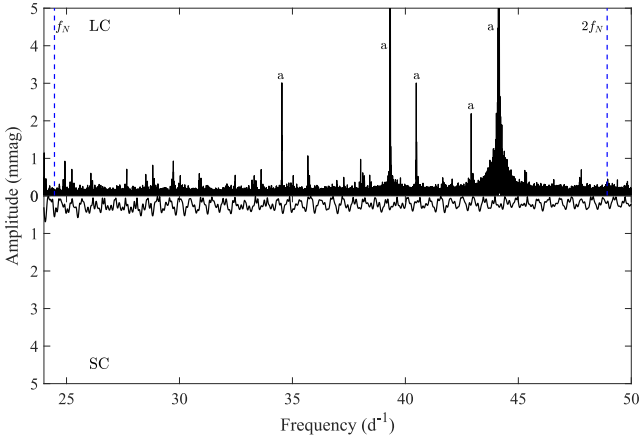

**Figure 6.** Comparison of amplitude spectra of the 4 yr LC and 9.7 day SC data in the super-Nyquist frequency range. Upper panel: amplitude spectra of the 4 yr LC; five alias frequencies are marked. Lower panel: amplitude spectra of the 9.7 day SC data.

Figure 8 shows the stellar evolutionary tracks with masses from 2.10 to 2.40 $M_\odot$ at five metallicity levels in the H-R diagram, as well as the position of KIC 3868420 constrained by effective temperature $T_{\rm eff} = 7800 \pm 100$ K from E. Niemczura et al. (2017) and the derived luminosity log $(L/L_\odot) = 1.815 \pm 0.023$. From this figure, it can be found that at different metallicity levels, the mass of KIC 3868420 can range from 2.16 to 2.30 $M_\odot$. All these models indicate that this star has evolved off the main sequence stage and is now traversing the Hertzsprung gap.

### 4.2. Pulsation Modeling with Rotation and Mixing Length

To refine the internal structure of the pulsator, we constructed an expanded grid of oscillation models using GYRE. Compared to the above analysis, the new grid explicitly incorporates rotation and a broader range of mixing-length parameters. Oscillation calculations were performed using the perturbative treatment for $p$-modes and the traditional approximation of rotation (TAR) for $g$-modes as implemented in GYRE v8.1 (R. H. D. Townsend & S. A. Teitler 2013). TAR provides a reliable description of the influence of the Coriolis force on high-order $g$-modes in moderately and slowly rotating stars and has become a standard tool in the interpretation of hybrid $p/g$ pulsators (C. Aerts et al. 2010; R.-M. Ouazzani et al. 2017).

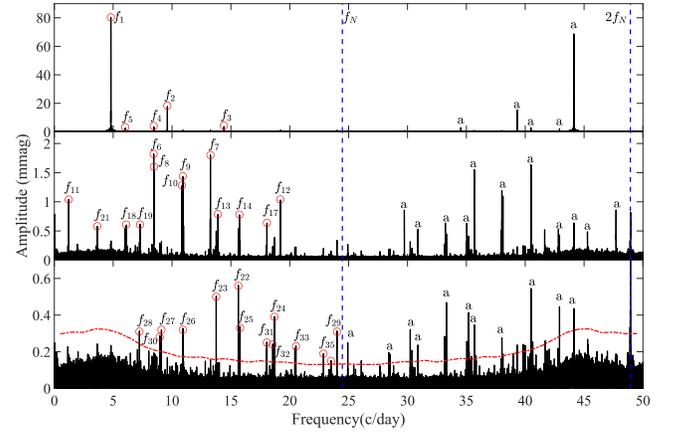

**Figure 7.** Top panel: amplitude spectrum of the 4 yr LC data beyond the LC Nyquist frequency ($f_N = 24.468$ day$^{-1}$, the vertical dashed lines), in which the first five strongest peaks are labeled, as well as the alias frequencies (labeled by "a"). Middle panel: amplitude spectrum after prewhitening five strongest peaks; another 16 significant peaks are labeled. For clarity, the frequencies $f_{15}$ (13.26020 day$^{-1}$) and $f_{16}$ (13.24660 day$^{-1}$), which are very close to $f_7$ (13.258368 day$^{-1}$), as well as $f_{20}$, have been omitted from the labels. Bottom panel: the frequencies $f_{22}$–$f_{36}$, as well as the residual spectrum after extraction of all significant frequencies. The red dashed line represents the threshold (S/N = 5.4) of significant frequency. Note that the frequencies $f_{34}$ (close to $f_{33}$) and $f_{36}$ (below $f_{35}$) are also not marked for the same reason.

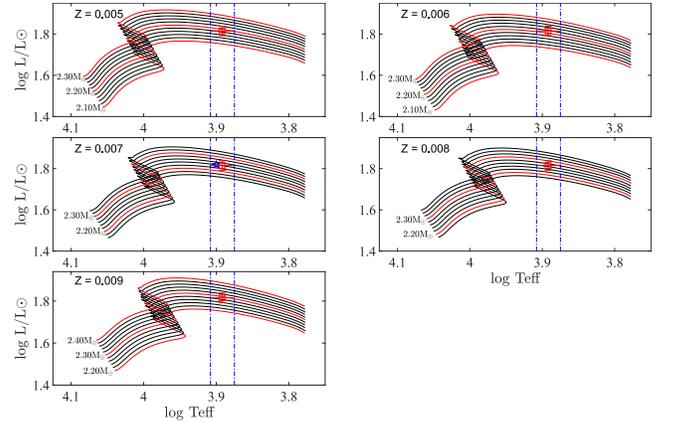

**Figure 8.** Stellar evolutionary tracks with different masses from the zero-age main sequence to post-main sequence. The values of metal abundance ($Z = 0.005$–0.009) are marked in the upper left corner of each panel. The red lines represent stellar models with masses of 2.10, 2.20, 2.30, and 2.40 $M_\odot$, respectively. The position of KIC 3868420 is constrained by effective temperature $T_{\rm eff} = 7800 \pm 100$ K from E. Niemczura et al. (2017) and the derived luminosity log $(L/L_\odot) = 1.815 \pm 0.023$. Blue dashed lines mark the $3\sigma$ range of $T_{\rm eff}$. The blue pentagrams mark the position of the best-fit model from a seismic model constrained by pulsational frequencies.

Guided by preliminary modeling that placed the star near $M \approx 2.20$–$2.30 M_\odot$, we calculated a set of evolutionary sequences spanning: $2.20 \leqslant M/M_\odot \leqslant 2.30$ in steps of 0.02, $Z = 0.005$–0.010, mixing-length parameter: $\alpha_{\rm MLT} = 1.5, 2.0, 2.5$, equatorial rotation velocities: $v_{\rm eq} = 10, 15, 20\,{\rm km\,s^{-1}}$. The adopted equatorial rotation velocities were chosen as an exploratory sensitivity grid motivated by the observed spectroscopic value $v \sin i = 5.7 \pm 1.2\,{\rm km\,s^{-1}}$ (Table 1). Because the inclination is unknown, this low projected velocity may correspond either to intrinsically slow rotation or to a more moderate rotator viewed at relatively low inclination. For representative inclinations $i \sim 90°$, $30°$, and $15°$, the implied equatorial velocities are approximately $v_{\rm eq} \sim 6, 11,$ and $22\,{\rm km\,s^{-1}}$, respectively. We therefore sampled slow-to-moderate rotation cases using $v_{\rm eq} = 10, 15,$ and $20\,{\rm km\,s^{-1}}$. These values should be regarded as an exploratory sensitivity test rather than an exhaustive determination of the star's true intrinsic rotation.

**Table 2**
Frequency, Amplitude, and Signal-to-noise Ratio of All the Significant Peaks Detected in KIC 3868420 Using Total LC, Sectors 0–5, and SC Data, Respectively

| ID $f_i$ | frequency-LC (day$^{-1}$) | Amplitude (mmag) | S/N | frequency-Sec (day$^{-1}$) | Amplitude (mmag) | S/N | frequency-SC | Amplitude | S/N | Comments |
|---|---|---|---|---|---|---|---|---|---|---|
| 1 | 4.8024429(2) | 81.149 | 1298.4 | 4.80244 | 80.871 | 682.0 | 4.80146 | 83.549 | 146.6 | Independent |
| 2 | 9.6048859(7) | 18.497 | 590.5 | 9.60489 | 18.475 | 360.8 | 9.60807 | 20.248 | 55.8 | 2f1 |
| 3 | 14.407325(3) | 4.475 | 188.1 | 14.40733 | 4.473 | 108.9 | 14.40953 | 5.527 | 60.7 | 3f1 |
| 4 | 8.455939(3) | 4.136 | 92.1 | 8.45596 | 3.474 | 58.9 | 8.46168 | 2.886 | 10.3 | Independent |
| 5 | 6.024730(4) | 3.371 | 63.1 | 6.02471 | 3.978 | 43.0 | 6.03010 | 4.567 | 23.9 | Independent |
| 6 | 8.444175(7) | 1.878 | 42.5 | 8.44421 | 2.349 | 37.2 | ⋯ | ⋯ | ⋯ | Independent |
| 7 | 13.258368(7) | 1.822 | 53.9 | 13.25827 | 1.649 | 29.0 | 13.26977 | 1.260 | 6.3 | f1+f4 |
| 8 | 8.457749(8) | 1.640 | 36.5 | 8.45811 | 1.698 | 26.7 | ⋯ | ⋯ | ⋯ | Independent |
| 9 | 10.921432(9) | 1.450 | 38.9 | 10.92124 | 1.572 | 28.3 | ⋯ | ⋯ | ⋯ | Independent |
| 10 | 10.82721(1) | 1.285 | 34.7 | 10.82711 | 1.491 | 26.2 | 10.85727 | 2.077 | 6.2 | f1+f5 |
| 11 | 1.22228(1) | 1.063 | 21.9 | 1.22215 | 1.304 | 13.3 | 1.21836 | 1.349 | 11.5 | f5-f1 |
| 12 | 19.20974(1) | 1.038 | 41.5 | 19.20974 | 1.032 | 28.4 | 19.21099 | 1.482 | 24.5 | 4f1 |
| 13 | 13.90163(2) | 0.802 | 26.5 | 13.90145 | 0.834 | 17.0 | 13.90722 | 1.245 | 8.0 | Independent |
| 14 | 15.72389(2) | 0.788 | 28.2 | 15.72367 | 0.804 | 18.5 | ⋯ | ⋯ | ⋯ | f1+f9 |
| 15 | 13.26020(2) | 0.729 | 21.6 | 13.26069 | 0.753 | 15.1 | ⋯ | ⋯ | ⋯ | f1+f8 |
| 16 | 13.24660(2) | 0.713 | 21.4 | 13.24672 | 0.913 | 18.1 | ⋯ | ⋯ | ⋯ | f1+f6 |
| 17 | 18.06080(2) | 0.642 | 25.7 | 18.06070 | 0.597 | 14.5 | 18.06877 | 0.466 | 4.0 | 2f1+f4 |
| 18 | 6.11898(2) | 0.604 | 11.3 | 6.11874 | 0.740 | 8.2 | ⋯ | ⋯ | ⋯ | f9-f1 |
| 19 | 7.28684(2) | 0.616 | 16.7 | 7.28665 | 0.644 | 9.3 | 7.27491 | 0.795 | 5.0 | Independent |
| 20 | 10.97643(2) | 0.603 | 16.3 | 10.97621 | 0.556 | 10.8 | ⋯ | ⋯ | ⋯ | Independent |
| 21 | 3.65356(2) | 0.582 | 10.3 | 3.65348 | 0.584 | 5.0 | ⋯ | ⋯ | ⋯ | f4-f1 |
| 22 | 15.62965(2) | 0.572 | 20.7 | 15.62956 | 0.666 | 16.6 | 15.65508 | 1.191 | 8.0 | 2f1+f5 |
| 23 | 13.75195(3) | 0.504 | 15.3 | 13.75216 | 0.560 | 11.6 | 13.74786 | 0.463 | 5.6 | Independent |
| 24 | 18.70413(3) | 0.390 | 15.4 | 18.70398 | 0.443 | 10.9 | 18.70868 | 0.602 | 4.3 | f1+f13 |
| 25 | 15.77886(4) | 0.340 | 12.3 | 15.77880 | 0.309 | 8.0 | ⋯ | ⋯ | ⋯ | f1+f20 |
| 26 | 10.91964(4) | 0.328 | 8.8 | ⋯ | ⋯ | ⋯ | ⋯ | ⋯ | ⋯ | Independent |
| 27 | 9.09920(4) | 0.331 | 10.0 | 9.09911 | 0.321 | 6.0 | ⋯ | ⋯ | ⋯ | f13-f1 |
| 28 | 7.23182(4) | 0.313 | 8.3 | 7.23149 | 0.254 | 3.8 | ⋯ | ⋯ | ⋯ | f7-f5 |
| 29 | 24.01219(4) | 0.307 | 13.1 | 24.01228 | 0.305 | 8.6 | 24.01003 | 0.351 | 9.4 | 5f1 |
| 30 | 8.94945(4) | 0.289 | 8.2 | 8.94945 | 0.334 | 5.8 | ⋯ | ⋯ | ⋯ | f23-f1 |
| 31 | 18.04907(5) | 0.255 | 10.3 | 18.04914 | 0.299 | 7.8 | ⋯ | ⋯ | ⋯ | 2f1+f6 |
| 32 | 18.55437(5) | 0.243 | 9.4 | 18.55447 | 0.259 | 6.7 | ⋯ | ⋯ | ⋯ | f1+f23 |
| 33 | 20.52629(5) | 0.233 | 9.6 | 20.52643 | 0.188 | 5.4 | ⋯ | ⋯ | ⋯ | 2f1+f9 |
| 34 | 20.43203(6) | 0.218 | 8.8 | 20.43188 | 0.262 | 7.2 | 20.47809 | 0.386 | 3.8 | 3f1+f5 |
| 35 | 22.86325(7) | 0.193 | 8.4 | 22.86274 | 0.171 | 5.0 | 22.90991 | 0.207 | 2.8 | 3f1+f4 |
| 36 | 23.50657(8) | 0.153 | 6.1 | 23.50620 | 0.148 | 4.3 | 23.50624 | 0.245 | 3.7 | 2f1+f13 |
| 37 | ⋯ | ⋯ | ⋯ | 35.04023 | 0.275 | 6.2 | ⋯ | ⋯ | ⋯ | ⋯ |
| 38 | ⋯ | ⋯ | ⋯ | 25.31578 | 0.208 | 3.9 | ⋯ | ⋯ | ⋯ | 4f1+f5 |

Each model was evolved from the ZAMS through the Hertzsprung gap until it cooled to $T_{\rm eff} = 6000$ K. The rotation rates were translated into angular velocities in GYRE, ensuring consistent structural and oscillation modeling.

For each stellar model, we computed adiabatic mode frequencies of degrees $l = 0, 1, 2$. The oscillation treatment included: low-order $p$-modes using first-order rotational perturbations, high-order $g$-modes using TAR (R. H. D. Townsend & S. A. Teitler 2013), and rotation-dependent frequency shifts that distinguish between different $v_{\rm eq}$. The TAR formalism is particularly suitable for high-order $g$-modes whose eigenfunctions probe the radiative interior, and has been validated extensively in the context of $\gamma$ Dor and slow pulsating B pulsators (C. Aerts et al. 2010; R.-M. Ouazzani et al. 2017). In total, we performed millions of pulsation calculations across the grid, and in each model calculation, the theoretical frequency was matched with the five observed frequencies. For each evolutionary track, only the model whose theoretical frequency is closest to the observed frequency will be selected.

For each model, we identified the nearest theoretical mode to each observed frequency $\nu_i^{\rm obs}$, forming the statistic:

$$\chi^2 = \frac{1}{N}\sum_{i=1}^{N}\left(\frac{\nu_i^{\rm obs} - \nu_i^{\rm grid}}{\sigma_{i,\rm eff}}\right)^2,$$
$$\sigma_{i,\rm eff}^2 = \sigma_{i,\rm obs}^2 + \sigma_{i,\rm grid}^2, \qquad (2)$$

where $\sigma_{i,\rm obs}$ and $\sigma_{i,\rm grid}$ represent the uncertainties in the observed and the grid-calculated frequencies, respectively.

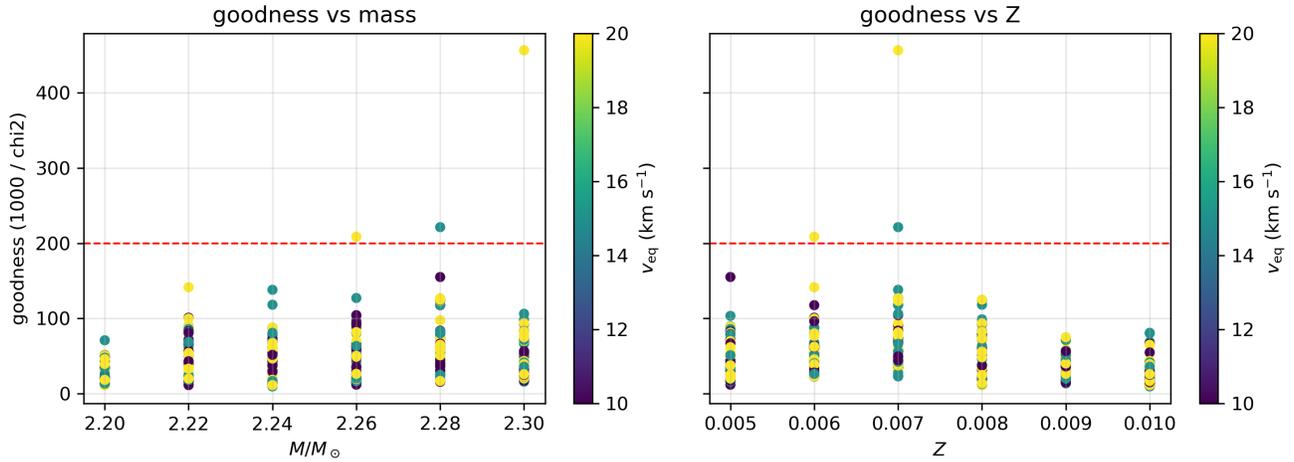

**Figure 9.** Goodness metric ($1000/\chi^2$) as a function of stellar mass (left) and metallicity (right) for all grid models. Color denotes the equatorial rotation velocity $v_{eq}$. The red horizontal dashed line marks goodness = 200, above which models are treated as acceptable solutions. The highest-goodness models near $M \approx 2.28\ M_\odot$ and $Z \approx 0.007$, indicating a narrow family of acceptable solutions within the restricted grid explored here, rather than a unique seismic determination.

To enhance interpretability across the grid, we define a goodness-of-fit metric as goodness = $1000/\chi^2$, where larger values indicate better agreement between the model and observations. The intrinsic precision of the theoretical frequencies in our grid calculations is limited to approximately 2%, a reflection of the star's rapid evolutionary stage, which leads to faster changes in stellar structure. In contrast, the observational frequency uncertainties are exceedingly small ($\sim 10^{-5}$) and are therefore negligible in our analysis. Based on this, we adopt a threshold of goodness = 200 to effectively separate acceptable from poor solutions, which is indicated by the dashed reference line in Figure 9.

Across the full grid, three models (see Table 3) yield the highest goodness values and define a narrow family of representative acceptable solutions within the explored model space: $M = 2.26$–$2.30\ M_\odot$, $Z = 0.006$–$0.007$, $\alpha_{MLT} = 1.5$–$1.8$, $v_{eq} = 15$–$20$ km s$^{-1}$, $T_{eff} = 7930$–$7980$ K, $\log(L/L_\odot) = 1.84$–$1.85$, $R \approx 4.41$–$4.43\ R_\odot$, and $\tau = (5.4$–$5.5) \times 10^8$ yr. These models support that KIC 3868420 is an evolved A-type pulsator crossing the Hertzsprung gap, although the true uncertainties are broader because of remaining degeneracies involving rotation, inclination, and mode identification.

The best-fitting models also provide candidate identifications for the five fitted frequencies. In the representative acceptable solutions, $f_1$ ($\approx 55.6\ \mu$Hz) and $f_{13}$ ($\approx 160.9\ \mu$Hz) are consistently associated with radial $p$-modes, while $f_4$, $f_5$, and $f_9$ are matched by $g$-mode candidates whose inferred radial orders vary with structure and rotation. These results support the interpretation of KIC 3868420 as a high-amplitude $\delta$ Sct–$\gamma$ Dor hybrid candidate, but the detailed mode labeling should not be regarded as unique.

Rotation modifies the oscillation spectrum through both the Coriolis force and structural effects. Under TAR, high-order $g$-modes experience frequency shifts proportional to the rotational frequency and mode geometry (R. H. D. Townsend & S. A. Teitler 2013), and these shifts become comparable to the spacing between adjacent high-order modes. Consequently, small changes in rotation can alter which theoretical mode best matches a particular observed frequency. This explains the modest but real degeneracy in the identification of $f_4$, $f_5$, and $f_9$.

In contrast, low-order $p$-modes are comparatively less affected across the slow-to-moderate rotation range explored here, so the $p$-mode interpretation of $f_1$ and the higher-frequency radial mode near $f_{13}$ remains more stable within the adopted grid. These different rotational sensitivities highlight the diagnostic power of hybrid pulsators: their $g$-modes probe the deep interior and rotation, while their $p$-modes constrain the outer envelope. The behavior of this star is fully consistent with the general theory of rotating pulsators (C. Aerts et al. 2010; R.-M. Ouazzani et al. 2017) and demonstrates the necessity of including rotation when interpreting mixed-mode frequency spectra in evolved A/F stars.

## 5. Discussion

Our frequency analysis reveals 36 significant peaks, including 11 independent modes spanning both the low- and high-frequency regimes. Together with the large photometric amplitude, this places KIC 3868420 among the small but growing group of evolved high-amplitude pulsators that show hybrid-like behavior. Forward modeling identifies a group of acceptable evolved models that place KIC 3868420 beyond the main sequence and support its interpretation as a high-amplitude $\delta$ Sct–$\gamma$ Dor hybrid candidate crossing the Hertzsprung gap. This is astrophysically important because the Hertzsprung gap is a short-lived evolutionary phase that remains sparsely represented in current asteroseismic samples.

At the same time, the detailed seismic solution is not unique. The observed spectroscopic value $v \sin i = 5.7 \pm 1.2$ km s$^{-1}$ suggests either intrinsically slow rotation or a more moderate rotator viewed at relatively low inclination. Because the inclination is unknown, $v \sin i$ alone cannot uniquely determine the true equatorial rotation rate, and the inferred rotational state therefore remains degenerate with mass and other structural parameters. Our adopted rotation grid should thus be viewed as an exploratory sensitivity test rather than a unique determination of the star's intrinsic rotation. In addition, no secure rotational multiplets are identified in the present data, and the azimuthal order $m$ is therefore unconstrained. Under these conditions, the mode assignments listed in Table 4, particularly for the $g$-mode candidates, should be regarded as representative identifications within the restricted model grid explored here rather than as unique seismic labels.

Table 3
Top Three Best-fitting Stellar Models from the Full Grid

| ID | $M/M_\odot$ | $Z$ | $v_{\rm eq}$ (km s$^{-1}$) | $T_{\rm eff}$ (K) | $\log(L/L_\odot)$ | $R/R_\odot$ | Age (yr) | Goodness |
|---|---|---|---|---|---|---|---|---|
| 1 | 2.30 | 0.007 | 20 | 7980 | 1.86 | 4.43 | $5.4 \times 10^8$ | 456.7 |
| 2 | 2.28 | 0.007 | 15 | 7930 | 1.84 | 4.41 | $5.5 \times 10^8$ | 221.1 |
| 3 | 2.26 | 0.006 | 20 | 7932 | 1.85 | 4.43 | $5.5 \times 10^8$ | 208.5 |

Table 4
Candidate Mode Assignments for Representative Acceptable Models

| $f_i$ | $\nu_{\rm obs}$ ($\mu$Hz) | $\nu_{\rm cal}$ ($\mu$Hz) | $\delta\nu$ | $(l, n)$ | Type |
|---|---|---|---|---|---|
| $f_1$ | 55.58347 | 55.31836 | 0.26511 | $(0, +1)$ | low-order radial p-mode |
| $f_4$ | 97.86904 | 98.09324 | 0.22420 | $(1, -36)$ or $(2, -66)$ | high-order g-mode |
| $f_5$ | 69.73023 | 69.24979 | 0.48044 | $(2, -96)$ or $(2, -97)$ | high-order g-mode |
| $f_9$ | 126.40465 | 126.39605 | 0.00860 | $(1, -26)$ or $(2, -48)$ | intermediate-order g-mode |
| $f_{13}$ | 160.89747 | 160.45485 | 0.44261 | $(0, +7)$ | higher-overtone radial p-mode |

**Note.** Radial orders are given in GYRE, with $n > 0$ for p-modes and $n < 0$ for g-modes. Frequencies are the five independent peaks identified in the observed amplitude spectrum. The listed $(l, n)$ values are candidate assignments obtained within the restricted model grid explored in this work. No secure rotational multiplets are identified, and the azimuthal order $m$ is not constrained by the present analysis. The assignments are therefore not unique, especially for the g-mode candidates.

This limitation is especially relevant because KIC 3868420 occupies an evolutionary regime in which seismic forward modeling is intrinsically difficult. Structural readjustment across the Hertzsprung gap proceeds on a much shorter timescale than main-sequence evolution, so the theoretical pulsation frequencies vary more rapidly with age and internal stratification than they do in more slowly evolving A/F stars. Rotation further increases this sensitivity, particularly for high-order g-modes. Within the TAR, modest changes in the assumed rotation rate can shift the candidate radial orders of the observed g-modes, while the low-order radial p-modes remain comparatively less affected (C. Aerts et al. 2010; R. H. D. Townsend & S. A. Teitler 2013; R.-M. Ouazzani et al. 2017). This behavior is consistent with our results, in which the candidate identifications of $f_4$, $f_5$, and $f_9$ vary across the grid, whereas the p-mode interpretation of $f_1$ and the higher-frequency radial mode near $f_{13}$ is more stable.

The broader astrophysical context also supports a cautious interpretation. Similar tensions between observed pulsation spectra and standard forward models have been reported for other complex evolved pulsators. For example, T.-Z. Yang et al. (2021) found that standard single-star models could not fully reproduce the oscillation properties of a hybrid pulsator in a binary system, while J.-S. Niu & H.-F. Xue (2025) reported anomalously large period-change behavior in the Hertzsprung-gap HADS star KIC 6382916. Although KIC 3868420 is not currently known to be a binary, these comparisons illustrate that rapid post-main-sequence evolution can make precise mode-by-mode matching especially sensitive to structural assumptions. Even so, the present analysis still supports a clear physical picture: KIC 3868420 is a rare evolved high-amplitude hybrid candidate showing coexisting p- and g-mode frequency ranges while traversing the Hertzsprung gap.

KIC 3868420 is therefore an interesting object for future follow-up. Additional constraints from higher-precision spectroscopy, rotational diagnostics from resolved mode patterns, or long-term monitoring of secular period changes would be particularly valuable for reducing the present degeneracies. Even without such constraints, the star already stands out as a rare example of hybrid-like pulsation persisting through the Hertzsprung-gap phase, which remains sparsely sampled in current asteroseismic studies.

## 6. Conclusion

We have carried out a photometric and exploratory asteroseismic study of KIC 3868420 using 4 yr of Kepler LC data, supplemented by SC observations for alias discrimination. The Fourier spectrum contains 36 significant frequencies, including 11 independent modes, and clearly shows the coexistence of low- and high-frequency pulsation regimes. Together with its large photometric amplitude, this places KIC 3868420 among the rare evolved high-amplitude pulsators exhibiting hybrid-like behavior.

Grid-based forward modeling with MESA and GYRE identifies a group of acceptable post-main-sequence models consistent with a star that has evolved beyond the TAMS and is now crossing the Hertzsprung gap. Within the restricted model grid explored here, representative solutions lie near $M \sim 2.26$–$2.30\,M_\odot$, $R \sim 4.41$–$4.43\,R_\odot$, and $\tau \sim 5.4$–$5.5 \times 10^8$ yr. These models support the interpretation of KIC 3868420 as an evolved high-amplitude $\delta$ Sct–$\gamma$ Dor hybrid candidate.

At the same time, the detailed seismic solution is not unique. The observed spectroscopic value $v \sin i = 5.7 \pm 1.2$ km s$^{-1}$ does not uniquely determine the equatorial rotation rate because the inclination is unknown, and no secure rotational multiplets are identified in the present data. As a result, the azimuthal order $m$ remains unconstrained tightly, and the individual mode assignments, especially for the g-mode candidates, should be regarded as tentative rather than unique.

KIC 3868420 is astrophysically valuable because the Hertzsprung gap is a rapid and sparsely sampled evolutionary phase in current asteroseismic studies. The detection of hybrid-like pulsation behavior in such an evolved high-amplitude star provides an informative observational case for testing pulsation physics during rapid post-main-sequence structural readjustment. Future progress will benefit from tighter spectroscopic constraints, more extensive exploration of rotational parameter space, and additional seismic diagnostics such as rotationally split mode patterns or long-term period-change measurements.

Looking ahead, a more exhaustive exploration of the parameter space, e.g., via Markov Chain Monte Carlo sampling, and forward modeling that incorporates more complete rotational physics (including second-order effects and, where possible, constraints on internal rotation) would be a natural extension of this work, but is beyond the scope of the present analysis. Expanding the sample of high-amplitude hybrids in short-lived evolutionary phases with ongoing and future space photometry (e.g., TESS and PLATO) will be essential for building an observationally grounded picture of pulsation and stellar structure across all evolutionary stages.

## Acknowledgments

This research is supported by the program of the National Natural Science Foundation of China (grant Nos. 12003020 and 12473043). We would like to thank the Kepler science team for providing such excellent data.

## ORCID iDs

Tao-Zhi Yang  https://orcid.org/0000-0002-1859-4949
Zhao-Yu Zuo  https://orcid.org/0000-0001-6693-586X